\documentstyle[12pt,epsfig,oldlfont]{article}
\newcommand{\npb}[3]{Nucl.~Phys.~#1 (19#2) #3}
\newcommand{\prl}[3]{Phys.~Rev.~Lett.~#1 (19#2) #3}

\newcommand{\pr}[3]{Phys.~Rev.~D#1 (19#2) #3}

\newcommand{\lsim}{\raisebox{-0.13cm}{~\shortstack{$<$ \\[-0.07cm] $\sim$}}~}

\newcommand{\ra}{\rightarrow}
\newcommand{\ee}{e^+e^-}
\newcommand{\s}{\\ \vspace*{-4mm} }
\newcommand{\nn}{\noindent}
\newcommand{\non}{\nonumber}
\newcommand{\beq}{\begin{eqnarray}}
\newcommand{\eeq}{\end{eqnarray}}

\newcommand{\tb}{\tan\beta}

\begin{document}

\begin{titlepage}

\begin{flushright}
KA--TP--30--1996\\
PM/96--39\\
December 1996 \\
\end{flushright}

\def\thefootnote{\fnsymbol{footnote}}

\vspace{1cm}

\begin{center}

{\large\sc {\bf The Higgs--photon--Z boson Coupling Revisited}}
\vspace{1cm}

{\sc A.~Djouadi$^{1,2}$, V. Driesen$^2$, W. Hollik$^2$ and A. Kraft$^2$} 

\vspace{1cm}

{\small $^1$ Physique Math\'ematique et Th\'eorique, UPRES--A 5032, \\
\vspace{0.1cm}
Universit\'e de Montpellier II, F--34095 Montpellier Cedex 5, France.}

\vspace*{4mm}

{\small $^2$ Institut f\"ur Theoretische Physik, Universit\"at Karlsruhe,\\
\vspace*{0.1cm}
D--76128 Karlsruhe, FR Germany.} 

\end{center}

\vspace{1.6cm}

\begin{abstract}

We analyze the coupling of CP-even and CP-odd Higgs bosons to a photon
and a Z boson in extensions of the Standard Model. In particular, we study in
detail the effect of charged Higgs bosons in two--Higgs doublet models, and the
contribution of SUSY particle loops in the minimal supersymmetric extension of
the Standard Model. The Higgs--$\gamma Z$ coupling can be measured in the decay
$Z \ra \gamma$+Higgs at $\ee$ colliders running on the $Z$ resonance, or in 
the reverse process Higgs $\ra Z \gamma$ with the Higgs boson produced at LHC. 
We show that a measurement of this coupling with a precision at the percent
level, which could be the case at future $\ee$ colliders, would allow to 
distinguish between the lightest SUSY and standard Higgs bosons in large 
areas of the parameter space. 

\end{abstract}

\end{titlepage}

\def\thefootnote{\arabic{footnote}}
\setcounter{footnote}{0}
\setcounter{page}{2}

\subsection*{1. Introduction}

The study of the electroweak symmetry breaking mechanism \cite{R1} is
one of the most important goals of present and future high--energy
colliders. Once the first evidence for Higgs particles is established,
it will be crucial to ascertain the underlying dynamics of the Higgs
sector. This can be achieved by measuring the couplings of the Higgs
particles to the other fundamental particles: in the Standard Model
(SM), fermions and gauge bosons acquire masses through the interaction
with the Higgs field and the size of their couplings are set by the
masses. This is a fundamental prediction of the Higgs mechanism which
has to be tested experimentally. \s 

Among these Higgs couplings, the couplings to photons are interesting in
many respects. First, the interaction of the Higgs particle with photons
does not occur at the tree--level since the photon is massless. The
Higgs--photon coupling is therefore induced by loops of heavy charged
particles. In the SM, this occurs via $W$ boson and heavy fermion
triangle loops. Since the couplings of these particles to the Higgs
boson grow with the mass, they balance the decrease of the triangle
amplitude with increasing loop mass, and the particles do not decouple
even for very large masses. Therefore these processes can serve to count
the number of heavy charged particles which couple to the Higgs boson.
\s 

In supersymmetric (SUSY) theories, the Higgs sector must be extended to
contain (at least) two doublets of scalar fields, leading to the existence 
of five
Higgs particles: two CP--even Higgs bosons $h$ and $H$, a CP--odd Higgs
boson $A$ as well as two charged Higgs particles $H^\pm$ \cite{R1}.
Depending on the value of $\tb$ [the ratio of the two vacuum expectation
values of the Higgs fields], the lightest SUSY Higgs boson $h$ is
constrained to be lighter than $M_h \lsim$ 80--130 GeV \cite{R2}
in the minimal version. For
small values $\tb \sim 1.5$, which are favored by Yukawa coupling
unification \cite{R3}, the Higgs boson has a mass which does not exceed
$M_h \sim 80$ GeV, and therefore can be produced at LEP2 \cite{R4}. In
the decoupling regime $M_A \sim M_H \sim M_{H^\pm} \gg M_Z$ \cite{R5},
the $h$ boson has practically the same properties as the SM Higgs
particle; the MSSM and SM Higgs sectors then look almost the same, and
are very difficult to be distinguished. However, additional
contributions to the Higgs--photon couplings will be induced by loops
with charged Higgs bosons, charginos and sfermions.
Since the SUSY particles do not couple to
the Higgs boson proportionally to their masses, their contributions
decouple for high masses.
 If, however, some of these
particles are not too heavy 
their contributions can be large enough to
allow for a discrimination between the lightest SUSY and standard Higgs
bosons even in the decoupling regime. \s 

The Higgs coupling to two photons has received much attention in the
literature \cite{R6,R7}. In the SM, the coupling of the Higgs boson
$H^0$ to a photon and a $Z$ boson has been discussed in Ref.~\cite{R8}.
If $M_Z < M_{H^0} \lsim 130$ GeV, the $H^0 Z \gamma$ vertex can be measured 
in the decay process $H^0 \ra Z\gamma$. At LHC the production rate for
light Higgs bosons is very large, $\sigma(gg \ra H^0) \sim 100$ pb
\cite{R9}, and despite of the small branching ratio BR($H^0 \ra Z
\gamma) \sim 10^{-3}$, one would still have ${\cal O}(10^{3})$ $Z
\gamma$ events if the luminosity is high enough, ${\cal L} \sim
10^{34}$cm$^{-2}$s$^{-1}$. The coupling could be therefore measured if
the background can be reduced to a manageable level and if the
theoretical prediction for the cross section is well under control
\cite{R10}. \s 

If $M_{H^0}<M_Z$, the $H^0 Z \gamma$ coupling can be measured in the
reverse decay $Z \ra H^0 \gamma$. At LEP and SLC, the rates are however
rather small, BR$(Z \ra H^0 \gamma) \lsim 10^{-5}$, leading to only a
few events. However, at future $\ee$ colliders \cite{R11} with the
expected integrated luminosities of $\int {\cal L} \sim  50$ fb$^{-1}$,
running a few weeks on the $Z$ resonance would allow to obtain a very
large sample of $Z \ra H^0 \gamma$ events. A very precise measurement of
the $H^0 Z \gamma$ coupling would be possible in this case. If the Higgs
boson is discovered at LEP2, one would then use the Next Linear $\ee$
Collider to measure the $H^0 Z \gamma$ coupling and check whether the
Higgs boson is SM--like or not. This measurement would be then
equivalent to measuring the $H^0 \gamma \gamma$ coupling at high--energy
$\gamma \gamma$ colliders \cite{R12}. \s 

In supersymmetric theories, the couplings of the light CP--even and
CP--odd Higgs bosons to $Z \gamma$ have been studied some time ago in
Ref.~\cite{R13} (see also \cite{R13a}). However, these analyses need to 
be updated for several reasons: (i) the radiative corrections in the MSSM 
Higgs sector turned 
out to be very large \cite{R2}, and therefore must be included; (ii) the
Higgs couplings to top squarks can be strongly enhanced if squark mixing
is included and this might induce large contributions to the Higgs--$Z
\gamma$ coupling, a possibility which has been overlooked; (iii)
stronger experimental bounds on the masses of charginos and sfermions
are now available \cite{R14}, eliminating a large part of the SUSY
parameter space where contributions from these particles are large; (iv)
finally, a fully analytic expression for the contributions of charginos
and top squarks with different masses is still lacking. \s 

In this paper, we address all the previous points. Furthermore, we
discuss in some details the possibility of using the Higgs--$Z
\gamma$ couplings to discriminate between the Standard Model and its
extensions. In particular, we analyze to what extent one can use
the SUSY loop contributions to distinguish between the standard and
light SUSY Higgs boson in the decoupling limit where all other Higgs
[and SUSY] particles are too heavy to be produced directly; we will show
that this is indeed possible if the coupling can be measured at the
percent level. We also show that in a general two Higgs--doublet
model, the contributions of charged Higgs bosons do not necessarily
decouple from the Higgs--$Z \gamma$ amplitude for large $H^\pm$ masses,
contrary to the SUSY case. \s 

The paper is organized as follows. In the next section, for
completeness and to set up the notation, 
we discuss the $H^0 Z \gamma$ coupling in the
Standard Model. In section 3, we analyse the Higgs--$Z\gamma$ coupling
in the two Higgs--doublet extension of the model. In section 4, we
discuss the various loop contributions of the SUSY particles to the
coupling in the MSSM, paying special attention to the small $\tb$
region and the decoupling limit. Our conclusions are given in
section 5.

\subsection*{2. The Higgs--Z--photon coupling in the SM}

For the sake of completeness and to fix our notation, we first 
discuss the Higgs--$Z \gamma$ coupling in the Standard Model. The 
$H^0 Z\gamma$ vertex is mediated by $W$ boson and heavy quark [in 
practice only top and bottom quark] loops; Fig.~1a. It can be decomposed 
into the following tensorial structure:
\begin{small}
\beq
V [Z^{\mu}(p_1),\gamma^{\nu}(p_2),H^0(p_3)]= F_0 p_2^{\mu} p_1^{\nu}
+ F_1 g^{\mu\nu} + F_2 p_2^{\mu} p_2^{\nu} + F_3 p_1^{\mu} p_2^{\nu} +
F_4 p_1^{\mu} p_1^{\nu} + F_5 \epsilon^{\mu\nu\alpha\beta} {p_1}_{\alpha} 
{p_2}_{\beta} \label{tdec}
\eeq
\end{small}
For on--shell particles, only the form factor $F_0$ contributes 
to the decay widths; normalized to $e^3/(s_W M_W)$ with $s_W^2=
1-c_W^2\equiv \sin^2\theta_W$, it is given by
\beq
F_0 = A_W+A_f \equiv  M_Z^2
\bigg[ \cot\theta_W F_{W} + \sum_{f}\, 2Q_f\,N_c\, \frac{m_f^2}{M_Z^2} \,
\frac{I^f_3-2s_W^2 Q_f}{s_W\,c_W}  \,F_f \bigg] \label{giz}
\eeq
with $Q_f$, $I^f_3$ and $m_f$ the charge, the weak isospin and the mass
of the fermion $f$; $N_c=1$ for leptons and $N_c=3$ for quarks. \s

In the following we discuss the various contributions in the case
of the decay $Z \ra H^0 \gamma$; the amplitudes for the reverse decay
can be simply obtained by crossing. In terms of the Passarino--Veltman 
three--point scalar functions \cite{R15}
\beq
C_{0,2}(m^2) \equiv C_{0,2}(M_Z^2,0,M_{H^0}^2;m,m,m)  
\eeq
 the fermionic and $W$
contributions\footnote{We have calculated the amplitudes in the Feynman 
gauge; however, the results for the fermion and $W$ loops are separately
gauge invariant if all external particles are on--shell.} are found to be 
\beq
F_f &=&   C_0(m_f^2) + 4 C_2(m_f^2) \non \\
F_W &=& 2 \left[ \frac{M_{H^0}^2}{M_W^2} (1-2 c_W^2)   + 2 ( 1-6 c_W^2 ) 
\right] C_2 (M_W^2) + 4 ( 1 - 4 c_W^2 ) C_0(M_W^2) 
\eeq
$C_0$ is the scalar integral, and $C_2$ is a short-hand notation for
$C_{2} \equiv C_{11}+C_{23}$ where the expressions of the $C_{ij}$ 
can be found in \cite{R16}.
Since there is only one mass running in the loops, the functions 
$C_0(m^2)$ and $C_2(m^2)$ have a rather simple form; in terms of the 
scaled variables $\tau_Z=4m^2/M_Z^2$ and $\tau_H=4m^2/M_{H^0}^2$, they
are given by the known expressions \cite{R1}
\beq
4 m^2 C_2(m^2) &=& \frac{\tau_Z \tau_H}{2(\tau_Z-\tau_H)}
+\frac{\tau_Z \tau_H^2}{2(\tau_Z-\tau_H)^2}
\Big( \tau_Z \left[f(\tau_Z)-f(\tau_H)\right] 
  + 2 \left[g(\tau_Z)-g(\tau_H)\right] \Big)
\non \\
4 m^2 C_0(m^2) &=& - \frac{2\tau_Z \tau_H}{\tau_Z-\tau_H}
\left[f(\tau_Z)-f(\tau_H)\right] 
\eeq
with the functions $f$ and $g$ defined by \cite{R1}
\begin{equation}
f(\tau) = \left\{ \begin{array}{ll}
{\rm arcsin}^2 \sqrt{1/\tau} & \tau \geq 1 \\
-\frac{1}{4} \left[ \log \frac{1 + \sqrt{1-\tau } }
{1 - \sqrt{1-\tau} } - i \pi \right]^2 \ \ \ & \tau <1
\end{array} \right.
\end{equation}
\begin{equation}
g(\tau) = \left\{ \begin{array}{ll}
\sqrt{\tau-1} \ {\rm arcsin} \sqrt{1/\tau} & \tau \geq 1 \\
\frac{1}{2} \sqrt{1-\tau} \left[ \log \frac{1 + \sqrt{1-\tau } }
{1 - \sqrt{1-\tau} } - i \pi \right] \ \ \ & \tau <1
\end{array} \right.
\end{equation}

The $W$ boson and top quark form factors 
$A_W$ and $A_f$ are shown in Fig.~2a as a function of
the Higgs boson mass.
The $b$ quark contribution as
well as the contributions of the other fermions are much smaller, due to
the small masses.
 In the range of interest 70 GeV $\lsim M_{H^0}
\lsim$ 130 GeV, the $W$ contribution is by far dominant, being one order
of magnitude larger than the top quark contribution; the two amplitudes
interfere destructively.  The QCD corrections to the top quark loop are
small, being of ${\cal O}(\alpha_s/\pi)$ \cite{R17}. \s 

The decay rate for the process $Z \ra H^0 \gamma$ reads in terms of $F_0$: 
\beq
\Gamma(Z \ra H^0 \gamma) = 
 \frac{\alpha\, G_F^2\, M_W^2 \,s_W^2 }{192\pi^4} 
\, M_Z^3 \left( 1 -\frac{M_{H^0}^2}{M_Z^2} \right)^3 |F_0|^2 \; .
\eeq
This rate, normalized to the total decay width $\Gamma_Z \simeq 2.5$ GeV, is
displayed in Fig.~2b. The branching ratio varies from $\sim 10^{-6}$ for
masses $M_{H^0} \sim 50$ GeV [which are ruled out in the SM, but are
still possible \cite{R14} in extensions of the model] to $\sim 10^{-7}$
for $M_{H^0} \sim 80$ GeV which can be probed at LEP2 \cite{R4}. This
means that only a few events can be produced at LEP1 with the present
sample of ${\cal O}(10^{7})$ $Z$ bosons collected by all  four
collaborations. However, a future collider with the expected yearly
integrated luminosity of $\int {\cal L} \sim 100$ fb$^{-1}$, will be able
to produce ${\cal O}(10^{10})$ $Z$ bosons per year; this translates into
${\cal O}(10^3)$ $Z \ra H^0 \gamma$ events for Higgs boson masses not too close to
the $M_{H^0} \sim M_Z$ threshold. Since the signal is
very clean [the photon being monochromatic and the decay products of the
Higgs boson, $H^0 \ra b\bar{b}$, being efficiently tagged with
micro--vertex detectors], one could measure the $H^0 Z \gamma$ coupling
with a statistical precision of a few percent allowing for a stringent
test of the $H^0Z \gamma$ coupling. One would therefore check whether
the coupling is SM--like, and measure with a good precision the $H^0WW$
and $H^0t\bar{t}$ coupling. \s 

If $M_{H^0} >M_Z$, the decay rate for the reverse process $H^0 \ra Z 
\gamma$ reads
\beq
\Gamma(H^0 \ra Z \gamma) = 
\frac{\alpha\, G_F^2 \,M_W^2\,s_W^2}{64\pi^4} 
\, M_{H^0}^3 \left( 1 -\frac{M_Z^2}{M_{H^0}^2} \right)^3 |F_0|^2 
\eeq
with $F_0$ given by eq.~(2). The branching
ratio when the $Z$ boson is decaying into electron and muon pairs  is
also shown in Fig.~2a [with
BR$(Z\ra \ee + \mu^+\mu^-) \sim 6\%$; other decays of the $Z$ bosons
will be rather difficult to extract from the background at the LHC].
 In the mass range $ M_{H^0} \sim
120$ GeV, the branching ratio is of the order of $10^{-4}$. With the
${\cal O}(10^6)$ Higgs bosons produced at LHC in the main production
mechanism $gg \ra H^0$ with an expected yearly luminosity of $\int {\cal
L} \sim 100$ fb$^{-1}$, a few hundred $H^0 \ra Z\gamma$ events could be
collected in a few years of running, if background problems can be
reduced to a manageable level. This rises the hope to measure the
$H^0 Z \gamma$ coupling once the Higgs boson is observed in the $H^0 \ra
\gamma \gamma$ mode for instance. 

\subsection*{3. The coupling in Two--Higgs Doublet Models}

In a Two Higgs--Doublet Model (THDM), the most general Higgs potential
compatible with gauge invariance, the correct breaking of the 
SU(2)$\times$U(1) symmetry and CP conservation is given by \cite{R1}
\begin{eqnarray}
V &=& \lambda_1 (|\phi_1|^2-v_1^2)^2 + \lambda_2(|\phi_2|^2-v_2^2)^2 
+\lambda_3[ (|\phi_1|^2-v_1^2)+(|\phi_2|^2-v_2^2) ]^2 \non \\
&& +\lambda_4[ |\phi_1|^2|\phi_2|^2-|\phi_1^\dagger \phi_2|^2 ] 
+\lambda_5[ \mbox{Re}(\phi_1^\dagger \phi_2)-v_1 v_2 ]^2 +\lambda_6[ 
\mbox{Im}(\phi_1^\dagger \phi_2) ]^2
\end{eqnarray}
with $\phi_1$, $\phi_2$ the two Higgs--doublet fields and $v_1,v_2$ their 
vacuum
expectation values. We have also assumed that the discrete symmetry $\phi_1
\to -\phi_1$ is only broken softly; an additional term, $\lambda_7 [ \mbox{Re}
(\phi_1^\dagger \phi_2)-v_1 v_2] \mbox{Im}(\phi_1^\dagger \phi_2)$, can be 
eliminated by redefining the phases of the scalar fields \cite{R1}. 
Parameterizing the Higgs doublets by
\beq
\phi_1={ \phi_1^+ \choose v_1+\eta_1+i\chi_1} \ , \ \ 
\phi_2={ \phi_2^+ \choose v_2+\eta_2+i\chi_2} \ \  
\eeq
one obtains for the mass terms in the CP--even Higgs sector 
\begin{equation}
(\eta_1,\eta_2) \left( \begin{array}{cc}
4(\lambda_1+\lambda_3)v_1^2+\lambda_5 v_2^2 & (4\lambda_3+\lambda_5)v_1 v_2 \\
(4\lambda_3+\lambda_5)v_1 v_2 & 4(\lambda_2+\lambda_3)v_2^2+\lambda_5 v_1^2
               \end{array} \right)
{ \eta_1 \choose \eta_2 }
\end{equation}
while in the CP--odd and charged Higgs sectors, one has
\begin{equation}
\lambda_6 (\chi_1,\chi_2)  \left( \begin{array}{cc} v_2^2 & -v_1 v_2 \\
-v_1 v_2  & v_1^2 \end{array} \right) { \chi_1 \choose \chi_2 } \ \ , 
\ \ \lambda_4 (\phi_1^-,\phi_2^-)  \left( \begin{array}{cc} v_2^2 & -v_1 v_2 \\
-v_1 v_2  & v_1^2 \end{array} \right) { \phi_1^+ \choose \phi_2^+ }
\end{equation}
Diagonalizing the mass matrices, one obtains the physical masses
\begin{eqnarray}
M^2_{H,h} = \frac{1}{2} \left[ {\cal M}_{11}+{\cal M}_{22} \pm
\sqrt{({\cal M}_{11}-{\cal M}_{22})^2+4{\cal M}_{12}^2}\ \right]  \non \\
M_A^2 = \lambda_6 v^2 \ \ \ \mbox{and} \ \ \ M_{H^\pm}^2 = \lambda_4 v^2
\hspace*{2cm}
\end{eqnarray}
with $v^2 \equiv v_1^2+v_2^2$~and ${\cal M}$ the mass matrix of eq.~(12). 
The mixing angle $\alpha$ in the CP--even Higgs sector is obtained from 
\begin{eqnarray}
\cos 2\alpha = \frac{{\cal M}_{11}-{\cal M}_{22}}%
{\sqrt{({\cal M}_{11}-{\cal M}_{22})^2+4{\cal M}_{12}^2}} \ , \ 
\sin2\alpha = \frac{2{\cal M}_{12}}%
{\sqrt{({\cal M}_{11}-{\cal M}_{22})^2+4{\cal M}_{12}^2}} 
\end{eqnarray}
Inverting these relations, one obtains the $\lambda$'s in terms of the
Higgs masses, and $\alpha, \beta$:
\begin{eqnarray}
\lambda_1 &=& 
\frac{1}{4\cos^2 \beta v^2}(\cos^2\alpha M_H^2+ \sin^2 \alpha M_h^2)
-\frac{\sin2\alpha}{\sin2\beta}\frac{M_H^2-M_h^2}{4v^2} +\frac{\lambda_5}
{4}(1-\frac{\sin^2\beta }{\cos^2\beta }) \ , \non \\
\lambda_2 &=& 
\frac{1}{4\sin^2 \beta v^2}(\sin^2\alpha M_H^2+ \cos^2 \alpha M_h^2)
-\frac{\sin2\alpha}{\sin2\beta}\frac{M_H^2-M_h^2}{4v^2} +\frac{\lambda_5}
{4}(1-\frac{\cos^2\beta }{\sin^2\beta }) \ , \non \\
\lambda_3 &=& \frac{\sin2\alpha}{\sin2\beta}\frac{M_H^2-M_h^2}{4v^2}
-\frac{\lambda_5}{4} \ \ , \ \ \lambda_4 = \frac{M_{H^\pm}^2} {v^2} 
\ \ , \ \ \lambda_6 = \frac{M_A^2}{v^2}
\end{eqnarray}
As one can see, the parameter $\lambda_5$ can not be fixed by the masses
and the mixing angles, unless one imposes a strict $\phi_1 \to -\phi_1$
symmetry resulting in $\lambda_5=0$, or 
by using the SUSY relation $\lambda_5=\lambda_6=M_A^2/v^2$,
as will be discussed later. \s 

In a general THDM, the four masses $M_{h}, M_H, M_A$ and $M_{H^\pm}$ as
well as the mixing angles $\alpha$ and $\beta$ are free parameters. The
interaction of the Higgs bosons with fermions are model--dependent; here,
we will consider the model where one Higgs doublet couples only to
up--type quarks, while the other doublet couples only to down--type
quarks and charged leptons [the so--called Model II \cite{R1} which
occurs in SUSY models for instance]. In this case, the 
couplings of the neutral Higgs
boson,  collectively denoted by $\Phi$, to fermions and massive
gauge bosons  are given in Tab.~1,
normalized to the SM Higgs couplings.
Due to CP invariance, the pseudoscalar $A$ does not couple to $W$ and $Z$
bosons. \s 

\begin{center}
\begin{tabular}{|c|c|c|c|c|} \hline
$\ \ \ \Phi \ \ \ $ &$ g_{\Phi \bar{u}u} $      & $ g_{\Phi \bar{d} d} $ &
$g_{ \Phi VV} $ \\ \hline
$h$  & \ $\; \cos\alpha/\sin\beta       \; $ \ & \ $ \; -\sin\alpha/
\cos\beta \; $ \ & \ $ \; \sin(\beta-\alpha) \; $ \ \\
 $H$  & \       $\; \sin\alpha/\sin\beta \; $ \ & \ $ \; \cos\alpha/
\cos\beta \; $ \ & \ $ \; \cos(\beta-\alpha) \; $ \ \\
$A$  & \ $\; 1/ \tb \; $\ & \ $ \; \tb \; $ \   & \ $ \; 0 \; $ \ \\ \hline
\end{tabular}
\end{center}
\vspace{0.2cm}
\nn {\bf Tab.~1} Higgs boson couplings to fermions and gauge bosons in 
the THDM compared to the SM Higgs couplings. 

\bigskip

For the \underline{CP--even Higgs bosons}, the couplings to photons and 
$Z$ bosons receive contributions from $W$ and top/bottom quark loops 
as well as contributions from charged Higgs boson loops (Fig.~1b). The 
structure of the vertex is again given by eq.~(1), and only the form factor 
$F_0$ contributes for the decay. It is given by [$\phi \equiv h,H$] 
\beq
F_0 &=& M_Z^2
\bigg[\cot\theta_W \,g_{\phi VV} F_{W}
 + \cot\theta_W \,g_{\phi H^+H^-}F_{H^\pm} \non\\
&& \hspace{1cm}+ \sum_{f}\, 2Q_f\,N_c\, \frac{m_f^2}{M_Z^2} \,
\frac{I^f_3-2s_W^2 Q_f}{s_Wc_W} \, g_{\phi ff} F_f
 \bigg] .
\eeq
The functions $F_W$ and $F_f$ are the same as previously, while the 
function $F_{H^\pm}$ for the charged Higgs contribution reads in term 
of the $C_2$ function defined previously (see also  \cite{R20}):
\beq
F_{H^\pm} &=& 4 C_{2}(M_{H^\pm}^2) .
\eeq
The couplings $g_{\phi VV}$ and $g_{\phi ff}$ can be taken from Tab.~1, while
the couplings of the CP--even neutral Higgs bosons to charged Higgs bosons 
in the THDM [using a normalization similar to the one for the $W$ boson and
the fermions] are found to be 
\begin{eqnarray} 
g_{hH^+H^-} &=&  \frac{M_h^2-\lambda_5 v^2}{M_W^2}\,
     \frac{\cos(\beta+\alpha)}{\sin2\beta}
       +\frac{2 M_{H^\pm}^2-M_h^2}{2M_W^2}\sin(\beta-\alpha) \non \\
g_{HH^+H^-} &=&  \frac{M_H^2-\lambda_5 v^2}{M_W^2}\,
      \frac{\sin(\beta+\alpha)}{\sin2\beta}
       +\frac{2 M_{H^\pm}^2-M_H^2}{2M_W^2}\cos(\beta-\alpha) \, . 
\end{eqnarray}
In the limit of very heavy $H^\pm$ bosons, the $C_2(M_{H^\pm}^2)$
function reduces to 
$$
C_2(M_{H^\pm}^2) \ra 1/(24 M_{H^\pm}^2) \ ,
$$
while the coupling of the $h$ boson [that we assume to be the lighter
CP--even Higgs state] to the charged Higgs boson approaches the limit  
$$
g_{h H^+H^-} \ra \frac{M_{H^\pm}^2}{M_W^2} \sin(\beta-\alpha) \, 
$$ 
assuming that $\lambda_5 v^2 \ll M_{H^\pm}^2$. This leads
to a final contribution which is proportional to $\sin ^2 (\beta-\alpha)
\equiv g_{hVV}^2$. Therefore, the charged Higgs contribution to the
$hZ\gamma$ coupling in a general THDM does not decouple, contrary to the
case of SUSY models as will be discussed later. However, the $H^\pm$
contribution is suppressed by the large factor $1/24$ and compared to
the $W$ boson loop [which in a THDM, is also damped by the factor $\sin
(\beta-\alpha)$ compared to the SM case], it is two orders of magnitude
smaller. The decay widths $\Gamma( Z \ra h\gamma)$ or $\Gamma(h \ra
Z\gamma$), given by eqs.~(8--9) with $M_{H^0} \ra M_{h,H}$, will
therefore be hardly sensitive to this loop effect. \s 

We now turn to the case of the \underline{pseudoscalar Higgs boson} $A$.
Due to CP--invariance, the $A Z \gamma$ coupling is induced only by
fermionic loops, since $A$ does not couple to $W$ and $H^\pm$ bosons.
Its tensorial structure is given by the same expression as in eq.~(1),
but here only the form factor $F_5$ contributes. The decay widths
$\Gamma(Z \ra A\gamma)$ or $\Gamma(A \ra Z \gamma)$ are given by
eqs.~(8--9) with $M_{H^0} \ra M_A$ and $F_0$ replaced by 
\beq
F_5 = - \sum_{f} \,  2Q_f\,N_c\, \frac{m_f^2}{M_W^2} 
(I^f_3-2s_W^2 Q_f)  g_{A ff} C_0(m_f^2) \, .
\eeq

\bigskip

In the general THDM, a numerical analysis is rather complicated [and not
very telling] since besides the four masses $M_h,M_H,M_A$ and
$M_{H^\pm}$, we have the mixing angles $\alpha$ and $\beta$ as
additional parameters, not to mention the parameter $\lambda_5$ which is
also independent. The Higgs--$Z \gamma$ couplings can vary widely
compared to the SM coupling, although the dominant $W$ boson amplitude
is always suppressed by the factors $\sin (\beta-\alpha)$ or
$\cos(\beta-\alpha)$ in the case of the CP--even bosons or absent in the
case of the pseudoscalar $A$. To simplify the discussion, we will use
the constraints provided by supersymmetry: in the MSSM, the Higgs sector
is described at the tree--level only by two free parameters that we
chose to be $\tb$ and the pseudoscalar mass $M_A$. The masses of the
CP--even Higgs bosons are given by 
\beq
M_{h,H}^2 =\frac{1}{2} \left[M_A^2+M_Z^2 \mp \sqrt{(M_A^2+M_Z^2)^2
-4M_A^2 M_Z^2 \cos^2 2\beta} \ \right]
\eeq
while the mass of the charged Higgs boson is simply given by 
\beq
M_{H^\pm}^2= M_A^2+M_W^2
\eeq
The mixing angle $\alpha$ is related to $M_A$ and $\tb$ by
\beq
\tan 2 \alpha = \tan 2\beta \frac{M_A^2+M_Z^2}{M_A^2-M_Z^2} \ \ , \ \ 
-\frac{\pi}{2} \leq \alpha \leq 0
\eeq
However, these relations are affected by large radiative corrections
\cite{R2} which must be taken into account. We will therefore include
the leading radiative correction to the Higgs masses and couplings which
grows as $m_t^4$ and logarithmically with the common squark mass that we
fix to 1 TeV, unless otherwise stated. In the MSSM, one has $1<
\tb<m_t/m_b$ from GUT restrictions, with the lower [$\tb \sim 1.6]$ and
the upper $[\tb \sim 50$] ranges favored by Yukawa coupling unification
\cite{R3}. We will mainly focus on the lightest CP--even Higgs boson, 
for which the maximum allowed value of the mass is about $M_{h}^{\rm max}
\simeq 80$ GeV for $\tb\sim 1.6$, and the particle is therefore accessible 
in $Z$ decays. In the high $\tb$ range, the maximal $h$ mass can reach 
values $M_h \sim 130$ GeV, and the decay $h \ra Z\gamma$ would be 
kinematically possible. \s

The $W$ boson amplitude $A_{W}=\cot\theta_W\,g_{hVV} F_W$ 
is shown in Fig.~3a as a
function of $M_h$ for the three values $\tb=1.6,5$ and 50. For low $h$
masses, $A_W$ is suppressed compared to the SM value, the suppression
being more effective with increasing $\tb$; in fact, for $\tb \sim 50$
the $W$ contribution almost vanishes. With increasing $h$ mass, $A_W$
approaches the SM value which is reached for $M_h =M_h^{\rm max}$. The
sum of the top and bottom loop contributions is also displayed in
Fig.~3a. Except when $M_h \sim M_{h}^{\rm max}$, where the the form
factor $A_f$ becomes SM--like, the $t$ contribution is suppressed by a
factor $g_{htt} \sim 1/\tb$, while the $b$ contribution is enhanced by
the factor $g_{hbb} \sim \tb$. Therefore, the $t$ contribution is
dominant for low $\tb$, while for large $\tb$ values the $b$
contribution [which has opposite sign compared to the SM case] is
strongly enhanced and becomes dominant. \s 

The contribution of the charged Higgs boson loop $A_{H^\pm}=\cot\theta_W
\, g_{hH^+H^-}F_{H^\pm}$ 
is shown in Fig.~3b as a function of $M_{H^\pm}$ for the
values $\tb=1.6,5$ and 50. The behavior can be understood by recalling
the expression of the $g_{hH^+H^-}$ coupling in the MSSM
\beq
g_{h H^+H^-} &=& \sin(\beta-\alpha) +\frac{\cos2 \beta
\sin(\beta+\alpha)}{2c_W^2}, 
\eeq
in which the
radiative correction must also be included; see for instance
Ref.~\cite{R7}.
For small $M_{H^\pm}$ implying small $M_h$, the coupling is strongly
suppressed for large $\tb$ values and the contribution $A_{H^\pm}$ is 
small. For $\tb \sim 1$, the suppression is rather mild and the $H^\pm$ 
contribution can be large, reaching a few percent of the $W$ contribution 
for $M_{H^\pm} \sim 100$ GeV. Contrary to the THDM, $A_{H^\pm}$ decreases 
with increasing $M_{H^\pm}$ since in the MSSM, the $g_{hH^+H^-}$ coupling 
does not scale like the charged Higgs mass, and the contribution is damped 
by a factor $1/M_{H^\pm} ^2$. The charged Higgs boson therefore yields small
contributions to the $hZ\gamma$ coupling and decouples from the vertex for 
high masses. 

\subsection*{4. The Higgs--Z--Photon coupling in the MSSM}

In the Minimal Supersymmetric extension of the Standard Model, the
couplings of the CP--even Higgs bosons to the photon and the $Z$ boson
receive, as in the Two Higgs--Doublet Model, contributions from $W$
bosons, top+bottom quarks and charged Higgs bosons. Extra contributions
also come from charged supersymmetric particles: sleptons, squarks
and charginos (Fig.~1c). The decay widths are again given by eqs.~(8--9)
with $M_{H^0} \ra M_\phi$, and the form factor $F_0$ reads: 
\beq
F_0 & = & M_Z^2
\bigg[ \cot\theta_W \,g_{\phi VV}F_{W}
+ \sum_{f}\, 2Q_f\, N_c\,\frac{m_f^2}{M_Z^2} \,
   \frac{I^f_3 - 2 s_W^2 Q_f}{s_Wc_W} \, g_{\phi ff} \,F_f \non \\
&&\hspace{1cm} 
 + \cot\theta_W \,g_{\phi H^+ H^-} F_{H^\pm}
 + \cot\theta_W F_{\chi^+} 
 + \sum_{\tilde{f}} N_c Q_{\tilde{f}} F_{\tilde{f}} \bigg] .
\eeq
The amplitudes from the fermions, $W$ and $H^\pm$ bosons are, as in 
the previous section, given by eqs.~(4) and (18), while the chargino 
contribution reads
\beq
F_{\chi^+} =  \sum_{j,k=1,2} \; \frac{m_{\chi_j^+}}{M_W}\, 
f\Big( m_{\chi_j^+},m_{\chi_k^+},m_{\chi_k^+} \Big)
\; \sum_{m,n=L,R}  g_{Z\chi^{+}_j\chi^{-}_k}^m  \,
\;g_{\phi\chi^{+}_k\chi^{-}_j}^n  \; .    \label{fichip} 
\eeq
The couplings of charginos to the $Z$ bosons are given by
\beq
g_{Z\chi^{+}_j\chi^{-}_k}^{L} = - \left(V_{i1} V_{j1}^*+ \frac{1}{2} V_{i2} 
V_{j2}^* - \delta_{ij} s_W^2 \right)  ,  \ \ 
g_{Z\chi^{+}_j\chi^{-}_k}^{R} = - \left( U_{i1} U_{j1}^* + \frac{1}{2} U_{i2} 
U_{j2}^* - \delta_{ij} s_W^2 \right) 
\eeq
while the couplings to the higgs bosons read 
\beq
g^L_{\phi \chi^+_i \chi_j^-} = Q^*_{ji} c_\phi - S^*_{ji} d_\phi \  ,\;\; 
g^R_{\phi \chi^+_i \chi_j^-} = Q_{ij} c_\phi - S_{ij} d_\phi 
\eeq
with $c_h/d_h= \sin \alpha/ \cos \alpha$ and $c_H/d_H = -\cos \alpha/ 
\sin \alpha$. The elements $Q_{ij}/S_{ij}$, as well as the matrices $V$ 
and $U$ which diagonalize the chargino mass matrix can be found in 
\cite{R18}. \s

The function $f$ entering the chargino form--factor is given by
\beq
f(m_1,m_2,m_2) & = & - 2 \Big[C_0(m_1,m_2,m_2)+ C_{1}(m_1,m_2,m_2) 
+ 2 C_{2}(m_1,m_2,m_2)  \non \\
 &  & \hspace{1.2cm} + 2C_{2}(m_2,m_1,m_1)-C_{1}(m_2,m_1,m_1) \Big]
\eeq
where $C_1$ and $C_2$ have now a more complicated structure since there are 
two particles with different masses inside the loop. In terms of the 
scalar Passarino--Veltman functions $A_0$, $B_0$ and $C_0$
one has
\beq
C_1(m_1,m_2,m_2) &\equiv & C_{11}(M_Z^2,0,M_h^2;m_1,m_2,m_2) \non \\
&=& \frac{B_0(M_h^2;m_1,m_2)-B_0(M_Z^2;m_1,m_2)}{M_Z^2-M_h^2}- 
C_0(M_Z^2,0,M_h^2;m_1,m_2,m_2) \non \\
C_2(m_1,m_2,m_2) &\equiv & C_{12}(M_Z^2,0,M_h^2;m_1,m_2,m_2) +
C_{23} (M_Z^2,0,M_h^2;m_1,m_2,m_2) \non \\
&=& \frac{m_1^2 - m_2^2 - M_Z^2 }{2(M_Z^2-M_h^2)^2}
\Big[ B_0(M_h^2;m_1,m_2) - B_0(M_Z^2;m_1,m_2) \Big] \non \\
&& + \frac{1}{2(M_Z^2-M_h^2)M_h^2 } \bigg[M_h^2 +2 m_2^2 M_h^2 
\,C_0 (M_Z^2,0,M_h^2;m_1,m_2,m_2) \non \\ 
&& + (m_2^2 -m_1^2) B_0(M_h^2;m_1,m_2) + A_0(m_1) - A_0(m_2) \bigg].
\eeq

The expressions of the scalar one--, two-- and three-- point functions 
$A_0, B_0$ and $C_0$ are 
\begin{equation}
A_0(m) = m^2 \left[1-\log \frac{m^2}{\mu^2} \right]
\end{equation}
\begin{eqnarray}
B_0(p^2,m_1,m_2) &=& 2 - \log \frac{m_1m_2}{\mu^2}
   +\frac{m_1^2-m_2^2}{p^2} \log\frac{m_2}{m_1} \\
& & +\frac{\lambda^{1/2}(p^2, m_1^2, m_2^2)}{p^2}
     \log\frac{m_1^2 + m_2^2 - p^2+ \lambda ^{1/2}(p^2, m_1^2, m_2^2)} 
{2m_1m_2} \nonumber
\end{eqnarray}
\begin{eqnarray}
C_0(M_2^2,0,M_1^2,m_1,m_2,m_2) =  \hspace*{8cm} \mbox{}\\
\frac{1}{M_1^2-M_2^2}  
\, \sum_{i=1}^2  \sum_{\sigma=\pm 1} (-1)^i  
\ {\rm Li}_2 \left[
\frac{2 M_i^2}{ m_2^2-m_1^2+M_i^2+\sigma\lambda^{1/2}(M_i^2,m_1^2,m_2^2)  
}\right] \nonumber
\end{eqnarray}
$\mu$ is the renormalization scale, and  the
ultraviolet poles in $A_0$ and $B_0$ are subtracted since the amplitudes 
are finite; $\lambda$ is the usual two--body phase space function:   
$\lambda(x,y,z)=x^2+y^2+z^2-2(xy+xz+yz)$.

Finally, the contribution of the squark and slepton loops to the 
$h Z\gamma$ couplings reads
\beq
F_{\tilde{f} } & = & -8 \sum_{j,k=1,2} \; g_{\phi\tilde{f}_j\tilde{f}_k} \; 
g_{Z\tilde{f}_k\tilde{f}_j} \; C_2(m_{\tilde{f}_j},m_{\tilde{f}_k},
m_{\tilde{f}_k})
\eeq
with the function $C_2$ defined in eq.~(30). 
The squark couplings to the 
$Z$ boson, including mixing between left-- and right--handed  sfermions, 
are given by
\beq
g_{Z \tilde{f}_1 \tilde{f}_1}  & = &
\frac{1}{s_W c_W} \left[ (I^f_3-Q_f s_W^2) \cos^2\theta_f 
-Q_f s_W^2 \sin^2\theta_f \right] \non\\
g_{Z \tilde{f}_2 \tilde{f}_2}  & = &
\frac{1}{s_W c_W} \left[ -Q_f s_W^2 \cos^2\theta_f 
+ (I^f_3-Q_f s_W^2) \sin^2\theta_f \right] \non\\
g_{Z \tilde{f}_1 \tilde{f}_2}  & = & \frac{-I_3^f}{s_W c_W}
\sin\theta_f\cos\theta_f
\eeq
The mixing is proportional to the fermion mass, and in practice is 
non--negligible only for the partners of the third generation fermions.

\smallskip
The couplings of the Higgs bosons to squarks have a more complicated 
structure because of the squark mixing. In the case of the light 
CP--even Higgs boson $h$, they read
\beq
g_{h \tilde{f}_1\tilde{f}_1} & = &
C_{LL}^h \cos^2\theta_f + C_{RR}^h \sin^2\theta_f
+2 C_{RL}^h \cos\theta_f\sin\theta_f \non \\
g_{h \tilde{f}_2\tilde{f}_2} & = &
C_{RR}^h \cos^2\theta_f + C_{LL}^h \sin^2\theta_f
- 2 C_{RL}^h \cos\theta_f\sin\theta_f \non \\
g_{h \tilde{f}_1\tilde{f}_2} & = &
C_{RL}^h (\cos^2\theta_f - \sin^2\theta_f)
+ (C_{RR}^h -C_{LL}^h ) \cos\theta_f \sin\theta_f
\eeq
with
\beq
 C_{LL}^h & = & (I_3^f-Q_f s_W^2)  g_{hVV} 
    - \frac{m_f^2}{M_Z^2} g_{hff} \non \\
 C_{RR}^h & = & (Q_f s_W^2) g_{hVV} 
    - \frac{m_f^2}{M_Z^2}  g_{hff}  \non \\
 C_{RL}^h & = - & \frac{m_f}{2 M_Z^2}
 \left[ A_f g_{hff}  - \mu g_{Hff} \right]
\eeq
where $A_f$ is the soft--SUSY breaking trilinear term and $\mu$ the 
Higgs--higgsino mass parameter; the couplings  $g_{hff}$ and $g_{\phi 
VV}$ are given in Tab.~1. The heavy CP--even Higgs boson couplings 
to squarks can be obtained from the previous ones, by performing the
substitutions $g_{h\dots}  \leftrightarrow g_{H\dots}$, $\sin \alpha 
\ra \cos \alpha$ and $\cos \alpha \ra \sin \alpha$. Note that in the case 
of the partners of the light fermions, the mixing angles and the fermion 
masses can be set to zero, and these couplings simplify considerably. \s

For the pseudoscalar Higgs boson $A$, only top+bottom quarks and the 
charginos are contributing to the $A Z \gamma$ amplitude because of
CP--invariance. The form factor $F_0$ has to be replaced by
\beq
F_5 = -  M_Z^2 \bigg[ \sum_{f} \, 2Q_f\,N_c\, \frac{m_f^2}{M_Z^2} \,
\frac{I^f_3-2s_W^2 Q_f}{s_Wc_W}  g_{A ff} C_0(m_f^2)
 - \cot\theta_W F_{\chi^+} \bigg] 
\eeq
with the chargino contribution
\beq
F_{\chi^+} = \sum_{j,k=1,2} \;\frac{m_{\chi_j^+}}{M_Z}\,
 g \Big( m_{\chi_j^+},m_{\chi_k^+},m_{\chi_k^+} \Big)\,
\Big(g_{Z\chi^{+}_j\chi^{-}_k}^R+g_{Z\chi^{+}_k\chi^{-}_j}^L\Big)
\Big(g_{A\chi^{+}_j\chi^{-}_k}^R-g_{A\chi^{+}_k\chi^{-}_j}^L\Big) . 
\eeq
The $g_{A \chi^+ \chi^-}$ couplings are given by
\beq
g^L_{\phi \chi^+_i \chi_j^-} = -Q^*_{ji} \sin\beta - S^*_{ji} \cos\beta \ , \ \
g^R_{\phi \chi^+_i \chi_j^-} = Q_{ij} \sin\beta + S_{ij} \cos\beta \ ,
\eeq
and the new function $g$ reads
\beq
g(m_1,m_2,m_2) & = & - 2 \Big[C_0(m_1,m_2,m_2)+ C_{1}(m_1,m_2,m_2)  
+ C_{1}(m_2,m_1,m_1) \Big] .
\eeq

In the MSSM, the CP--even Higgs boson $H$ is always heavier than $M_Z$
and therefore only the decay $H \ra Z\gamma$ is kinematically possible.
However, even before allowing the $Z$ boson to decay into charged
leptons, the branching ratio BR$(H\ra \gamma Z)$ is very small and the
process will be very difficult if not impossible to be seen at the LHC.
This is also the case of the decay $A \ra Z\gamma$. The process $Z\ra A
\gamma$ will be possible if $M_A \lsim 80$ GeV, but then the
pseudoscalar Higgs boson can be discovered at LEP2 in the associated
production mechanism $\ee \ra hA$ \cite{R4} and its properties can be
studied. In the numerical analysis, we therefore focus on the light
CP--even Higgs boson $h$ and study in particular the low $\tb$ scenario
in which $M_h \lsim 80$ GeV and the $h$ boson can be produced at LEP2 in
the process $\ee \ra hZ$. We will pay special attention to the
decoupling limit where the $h$ boson mimics the SM Higgs particle
and the measurement of the $hZ\gamma$ coupling at future $\ee$ linear
colliders running at the $Z$ resonance could help discriminating between
the SM and MSSM scenarios. \s 

The contributions of the $W$, SM fermion and charged Higgs boson loops
to the $hZ\gamma$ have already been discussed in the THDM with the MSSM
constraints. As discussed previously, for large values of $M_A$, the $W$
and quark contributions are as in the SM, while the charged Higgs boson
decouples and its contribution is negligible. The contributions of the
slepton and the scalar partners of the light quarks, neglecting sfermion 
mixing, are shown in Fig.~4 as functions of the
masses and for the three values $\tb=1.6, \, 5$ and 50 with $M_A$ fixed
to 1 TeV. We have summed over all slepton and squark [except stop]
contributions, and used common masses $m_{\tilde{l}}$ and
$m_{\tilde{q}}$. 
As in the case of the charged Higgs boson, slepton and squark loop
contributions to the $h \ra Z \gamma$ decay width are very small, except
when these particles are very close to their allowed mass values
\cite{R14}. For loop masses above 150 GeV, the current experimental
bound on squark masses, they do not exceed the level of a few permille
of the dominant $W$ contribution and will therefore hardly
be detected. \s 

The contribution of the top squark loops to the $h \gamma Z$ vertex
depend on the soft SUSY--breaking scalar masses $m_{\tilde{t}_L}$ and 
$m_{\tilde{t}_R}$ [which are taken in general to be equal], as well as
on the soft--SUSY breaking trilinear term $A_t$, the Higgs--higgsino mass 
parameter $\mu$ and $\tb$. These parameters also determine the masses and 
mixing angle of the scalar top quarks and their couplings to the Higgs 
bosons. In the decoupling limit, $M_A \gg M_Z$ the contribution  
only depends on the combination $m_{t}^{LR} =A_t -\mu/\tb$ and 
$m_{\tilde{t}_L}$ which can be traded against the mass of the lightest 
top squark $m_{\tilde{t}_1}$. In Fig.~5a, we show contour plots in the 
$(m^{LR}_t,m_{\tilde{t}_1})$ plane for
which the contribution $A_{\tilde{t}}$, which includes the amplitudes of both
top squarks, is $0.5,0.2$ and $0.1$. For large $M_A$, i.e. in the decoupling 
limit, the amplitude $A_{\tilde{t}}$ is symmetric for positive and negative 
$m^{LR}_t$ values. For large $|m^{LR}_t|$, the contributions are large and 
negative; for light enough top squarks, $m_{\tilde{t}_1} \sim 150$ GeV, they 
can reach the value $A_{\tilde{t}} \sim -.5$ for $|m^{LR}_t| \sim 1$ TeV, 
i.e. at the level of the top quark contribution. For a given $m_{\tilde{t}_1}$,
$A_{\tilde{t}}$ is larger for higher values of $m^{LR}_t$, because in this 
case the coupling $h \tilde{t} \tilde{t} \sim m^{LR}_t$ is strongly enhanced. 
For large $m_{\tilde{t}_1}$, the two top squarks will have comparable masses 
and their amplitudes will partly cancel each other, leading to the contour
$A_{\tilde{t}}=0$. For small $|m^{LR}_t|$, there is a region around $m^{LR}_t
=0$ where no solution for $m_{\tilde{t}_1} <m_t$ is allowed when diagonalizing 
the mass matrix and this region is already excluded by CDF/D0 data since there 
one has $m_{\tilde{q}}<150$ GeV \cite{R14}. The amplitudes in this region are
positive and can reach values $A_{\tilde{t}}=0.2$ which decrease with
increasing top squark mass as expected. \s

Fig.~5b shows the contribution $A_{\tilde{t}}$ for $M_A =100$ GeV, i.e.
away from the decoupling limit, and $\tb=1.6$ [solid lines] and 50
[dashed lines]. For low $\tb$ values, the symmetry around $m^{LR}_t=0$
is lost; the picture is the same as in Fig.~5a for positive $m_t^{LR}$
values, but the contribution $A_{\tilde{t}}$ becomes smaller for
negative $m_{t}^{LR}$. For the high $\tb$ scenario, the contribution
$A_{\tilde{t}}$ becomes very small. \s 

The contribution of the charginos to the $h Z \gamma$ coupling depends 
on $\tb$, the gaugino mass parameter $M_2$ and the Higgs--higgsino mass 
parameter $\mu$ [these parameters also fix the chargino masses]. 
The form factor $A_\chi $ is shown in Fig.~6 in the $(M_2,\mu)$ plane for 
$\tb=1.6$ and $M_A=1$ TeV and $M_A=80$ GeV. Contours for $|A_\chi| =2,1,0.5
$ and $0.2$ as well as the region of the parameter space for which 
the lightest chargino mass is larger than 70 GeV [which approximately
corresponds to the current experimental limit] have been drawn. 
The chargino contributions are rather large close to the $M_{\chi^+}
=70$ GeV boundary, reaching values $A_{\chi^+} \sim 1$, and become smaller 
when one moves away from this boundary. However, in a large part of the 
$(M_2,\mu)$ parameter space, the chargino contribution is larger than 
$A_{\chi^+} =0.2$ and does not strongly depend on $M_A$. In fact, in this 
case, $A_{\chi^+}>0.1$ in the entire parameter space $M_2, \mu <500$ GeV
which leads to a deviation of the $h \ra Z\gamma$ coupling by more than
one percent.

\subsection*{5. Summary} 

We have analyzed the Higgs--$Z\gamma$ coupling in the Minimal Supersymmetric 
extension of the Standard Model. We have included the large radiative 
corrections in the Higgs sector, updated the contributions of the charginos 
and the top squarks to the coupling, and given fully analytic expressions for 
these contributions. In Two Higgs--Doublet Models, we have shown that the 
additional contribution from charged Higgs bosons do not necessarily decouple 
from the amplitude, if the $H^\pm$ mass is large. \s

We have payed special attention to the case of the MSSM lightest CP--even 
Higgs boson $h$ in the decoupling limit, where it has almost exactly the 
properties of the standard Higgs particle. The contributions of the
$W$ and top quark loops to the $hZ \gamma$ coupling are the same as in the 
SM, but additional contributions are induced by chargino and top 
squark loops. In the low $\tb$ scenario where the $h$ boson is lighter than 
$\sim 80$ GeV, these contributions can induce large deviations of the $Z \ra
h \gamma$ decay width from the SM value, even in the decoupling limit. 
This is illustrated in Fig.~7, where the deviations due to charginos and 
$\tilde{t}$ quarks are shown for two values of the masses $100$ and $250$ 
GeV. As can be seen, in very large areas of the MSSM parameter space, the 
deviations can exceed the level of several percent. \s

If future $\ee$ linear colliders could spend a few months running at the 
$Z$ resonance, a large sample of $Z \ra \gamma$+Higgs decays could be collected 
with the expected high--luminosities, if the decay is kinematically allowed. 
In this case, the Higgs--$Z \gamma$ coupling could be measured with a precision at the 
percent level, allowing a stringent test of the coupling. In the Standard 
Model, the couplings of the Higgs particle to $W$ bosons and top quarks can 
be measured with a good precision. In the MSSM, since the contributions 
of the genuine SUSY particles to the decay width exceed the percent 
level in large areas of the parameter space, the $h$ boson can be 
distinguished from the standard Higgs boson even in the  decoupling 
limit. The measurement of the Higgs--$Z\gamma$ coupling in the decay $Z \ra 
\gamma+$ Higgs will be in this sense, competitive with the measurement
of the Higgs couplings to two photons at high--energy $\gamma \gamma$ 
colliders.

\bigskip
\nn {\bf Acknowledgement:} Discussions with J. Rosiek are gratefully
acknowledged. 

\newpage

\def\npb#1#2#3{{\rm Nucl. Phys. }{\rm B #1} (#2) #3}
\def\plb#1#2#3{{\rm Phys. Lett. }{\rm #1 B} (#2) #3}
\def\prd#1#2#3{{\rm Phys. Rev. }{\rm D #1} (#2) #3}
\def\prl#1#2#3{{\rm Phys. Rev. Lett. }{\rm #1} (#2) #3}
\def\prc#1#2#3{{\rm Phys. Rep. }{\rm C #1} (#2) #3}
\def\pr#1#2#3{{\rm Phys. Rep. }{\rm #1} (#2) #3}
\def\zpc#1#2#3{{\rm Z. Phys. }{\rm C #1} (#2) #3}
\def\nca#1#2#3{{\it Nuevo~Cim.~}{\bf #1A} (#2) #3}
%

\newpage

\subsection*{Figure Captions}

\begin{itemize}

\item[{\bf Fig.~1:}]
Feynman diagrams contributing to the Higgs coupling to a photon and 
a $Z$ boson in the SM (a), additional contributions in the THDM (b)
and additional contributions in the MSSM(c). 

\item[{\bf Fig.~2:}]
The $W$ and top quark form factors in the SM as a function of the Higgs 
boson mass (a) and the branching ratios of the decays $H^0 \ra Z \gamma \ra 
l^+ l^- \gamma$ and $Z \ra H^0 \gamma$ in the SM. 

\item[{\bf Fig.~3:}]
The amplitudes for the contribution of the $W$ boson loop and of the sum 
of the $t,b$ loops as a function of $M_h$ (a) and the contribution of
the charged Higgs boson loops as a function of $M_{H^\pm}$ (b). The 
contributions are in a THDM with $\tb=1.6,5$ and 50. 

\item[{\bf Fig.~4:}] 
The amplitudes for the contribution of the slepton (a) and squark (except 
stop) loops (b) as functions of the loop masses for $\tb=1.6, 5$ and 50. 
We have neglected sfermion mixing and have set $M_A=1$ TeV.  

\item[{\bf Fig.~5:}]
Contours in the ($m^{LR}_t, m_{\tilde{t}_1})$ plane, for which the
contribution of the top squark loops to the $hZ \gamma$ coupling is 
$|A_{\tilde{t}}| =0.5, 0.2$ and $0.1$ for $M_A=1$ TeV (a) and $M_A=80$
GeV (b) with $\tb=1.6$.

\item[{\bf Fig.~6:}]
Contours in the ($M_2, \mu)$ plane for $\tb=1.6$ and $M_A=1$ TeV (a)
and $M_A=80$ GeV (b)
for which the contribution of the chargino loops to the $hZ \gamma$
coupling is $|A_{\chi}|=0.2, \ 0.5\ , 1$ and $2$. Also included 
are the contours for which the lightest chargino mass is $m_{\chi_1^+}
=70$.

\item[{\bf Fig.~7:}]
The deviations of the $h \ra Z\gamma$ decay width from the Standard 
Model value [in \%] for $\tb=1.6$ and the loop
masses $m_i =100$ and $250$ GeV. (a) Deviations due to the chargino 
loops as a function of $M_2$ for both signs of $\mu$, and (b) deviations 
due to the top squark loops as a function of $m_{t}^{LR}$. 

\end{itemize}

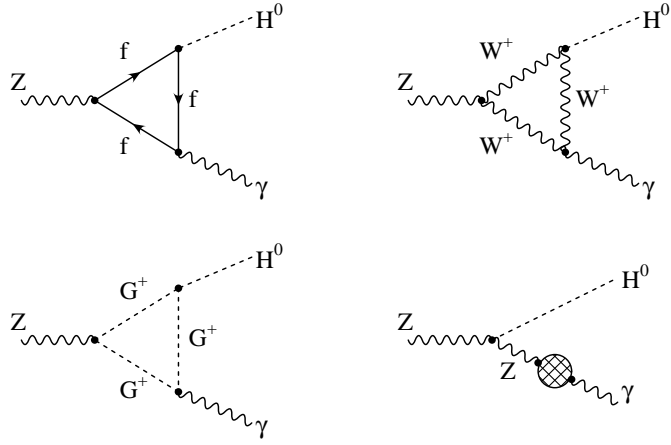
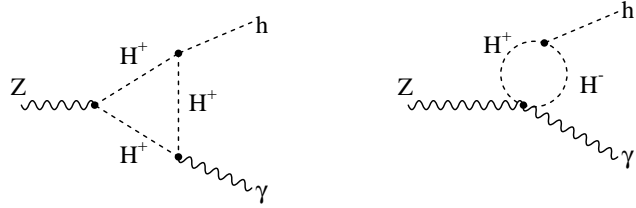
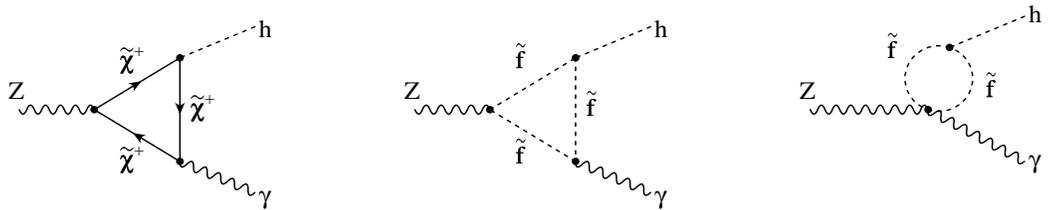
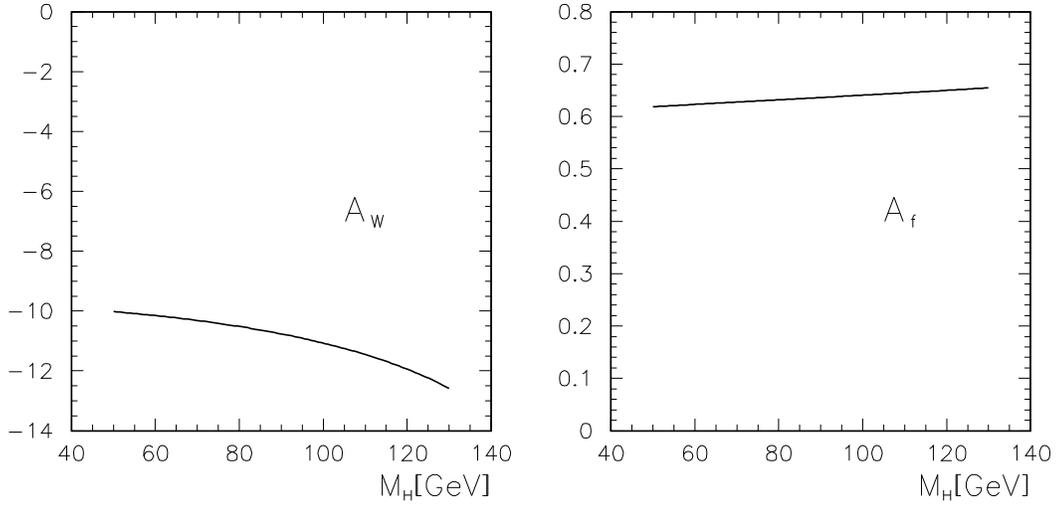
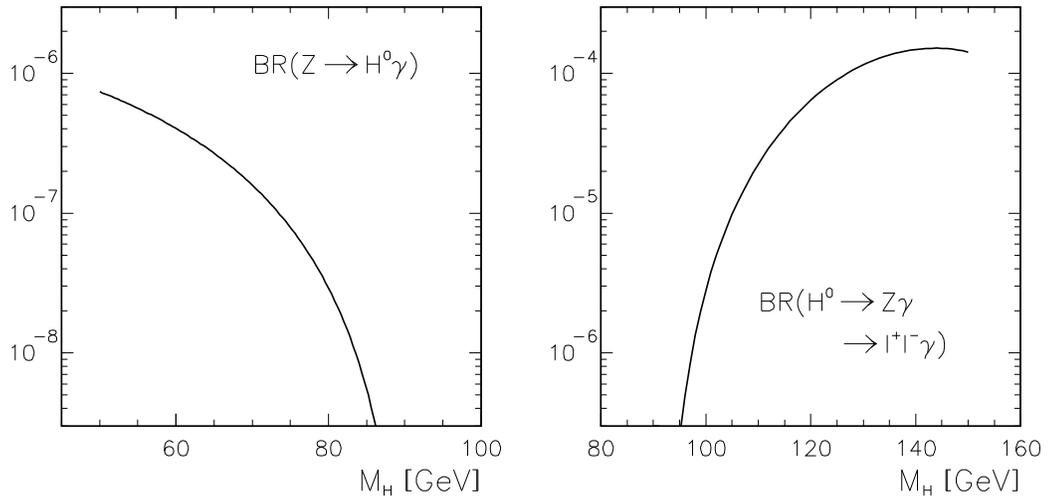
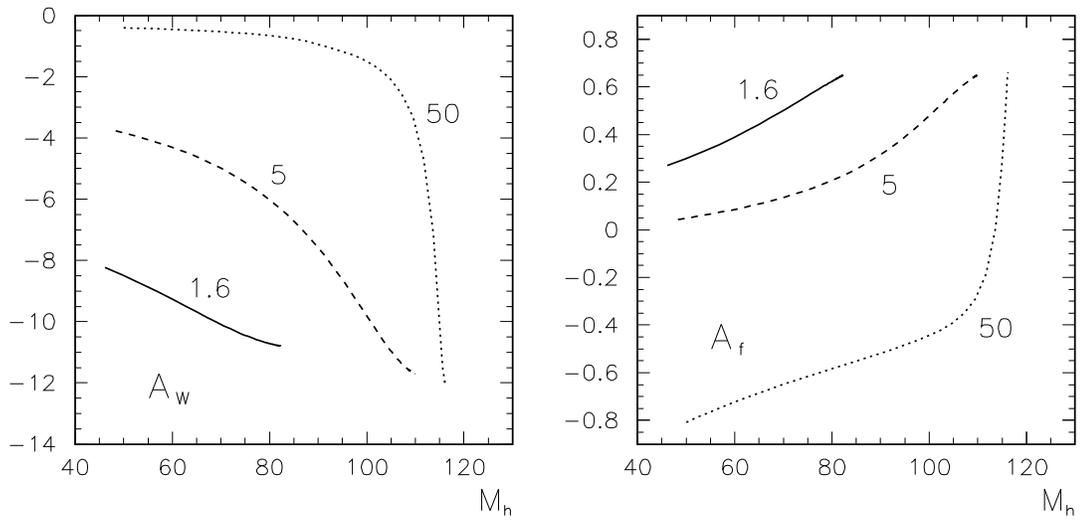
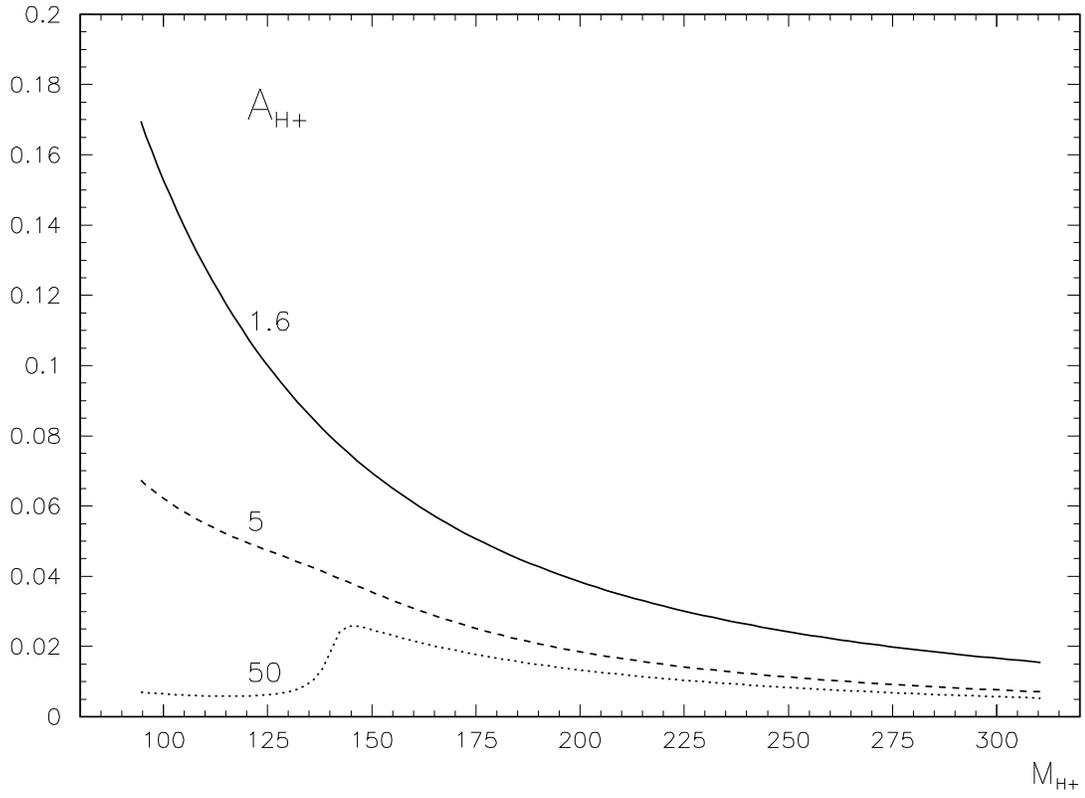
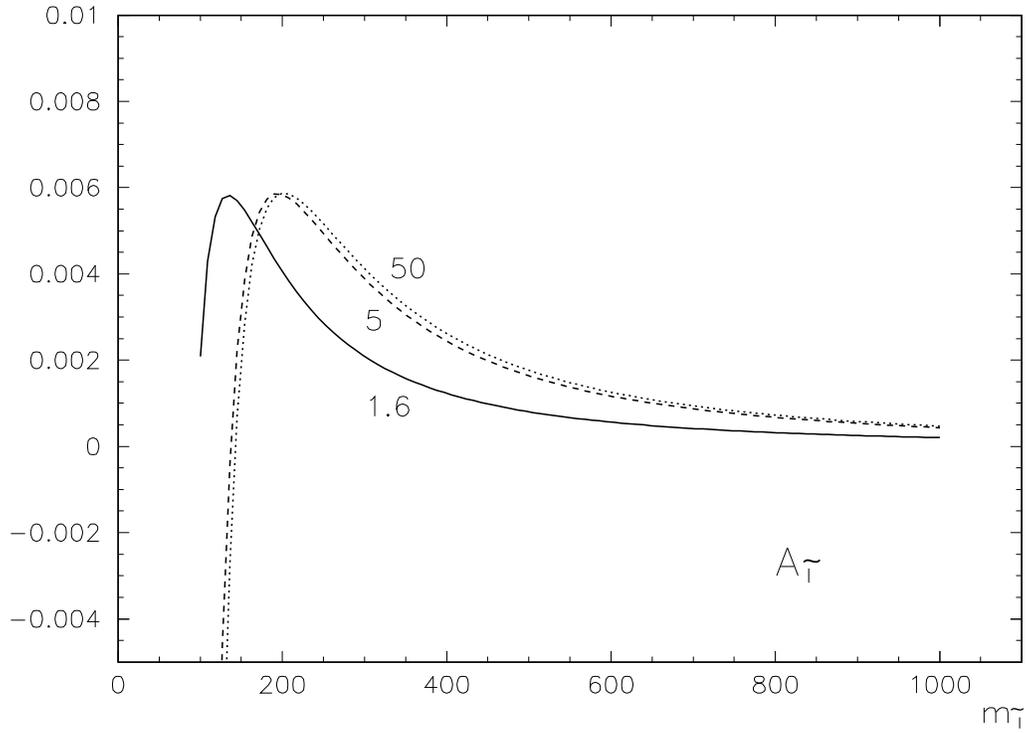
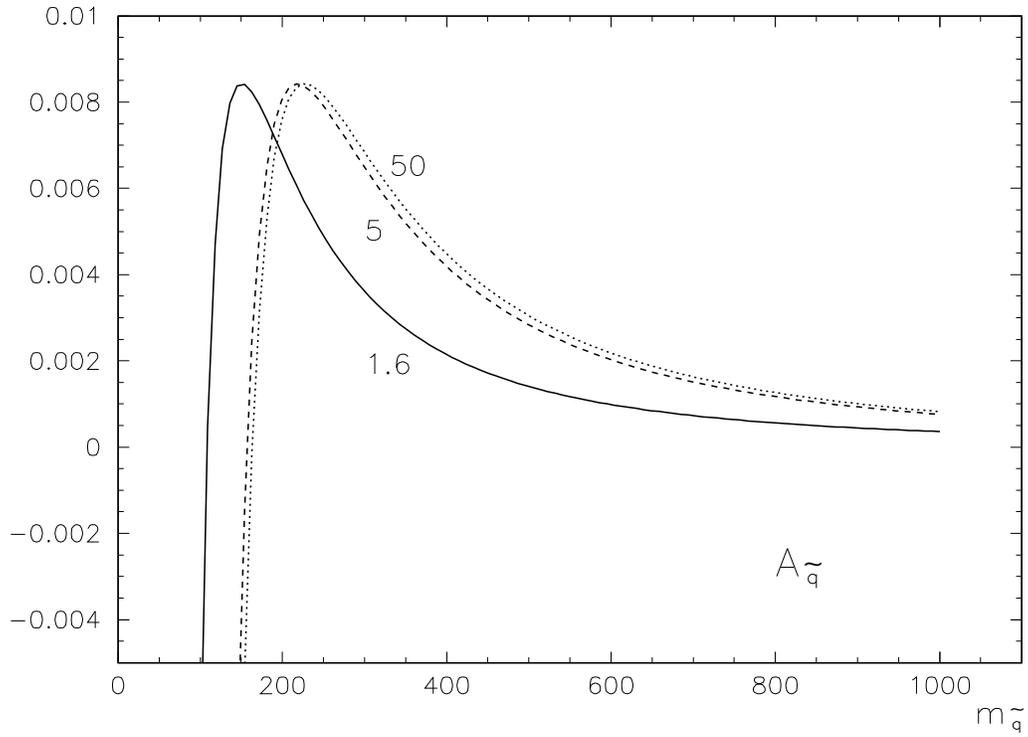
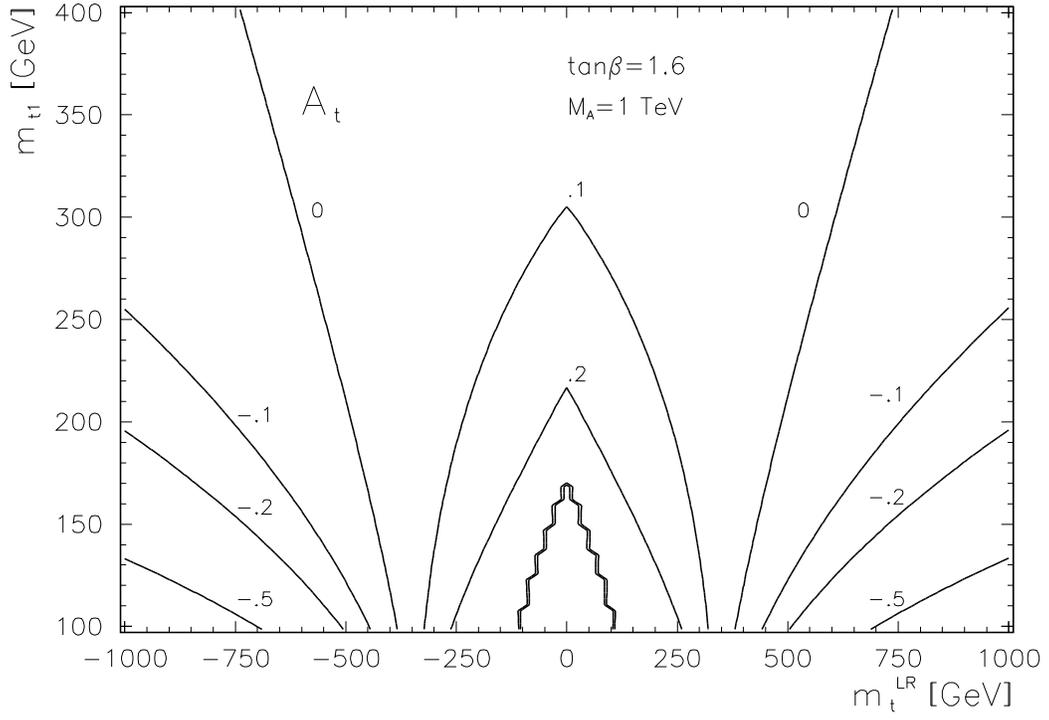
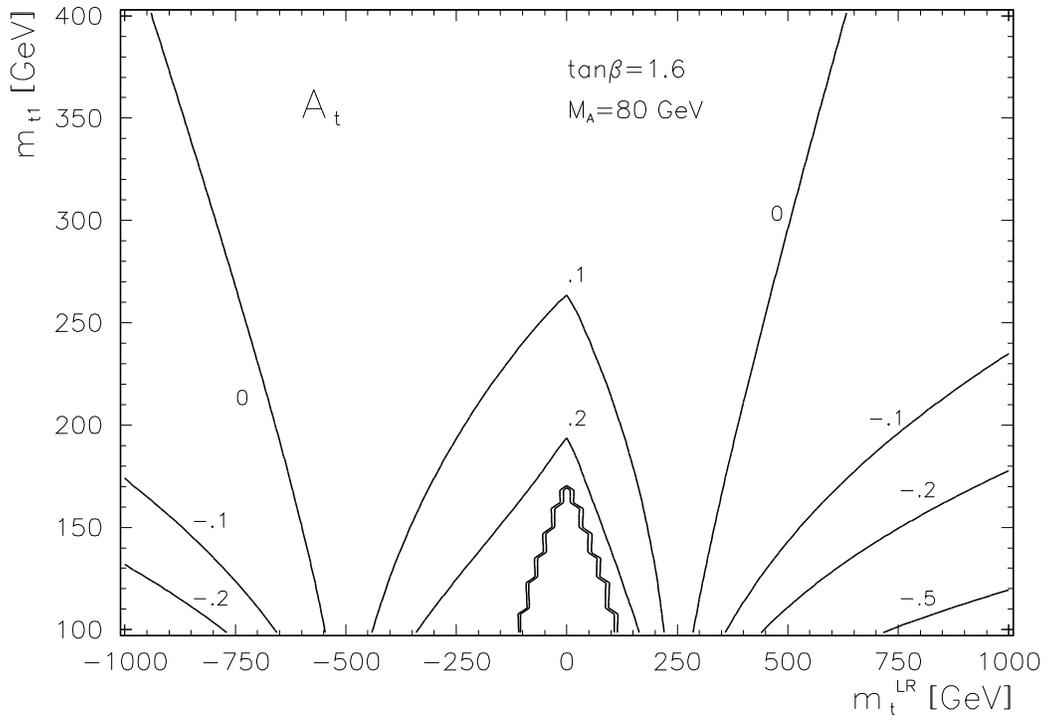
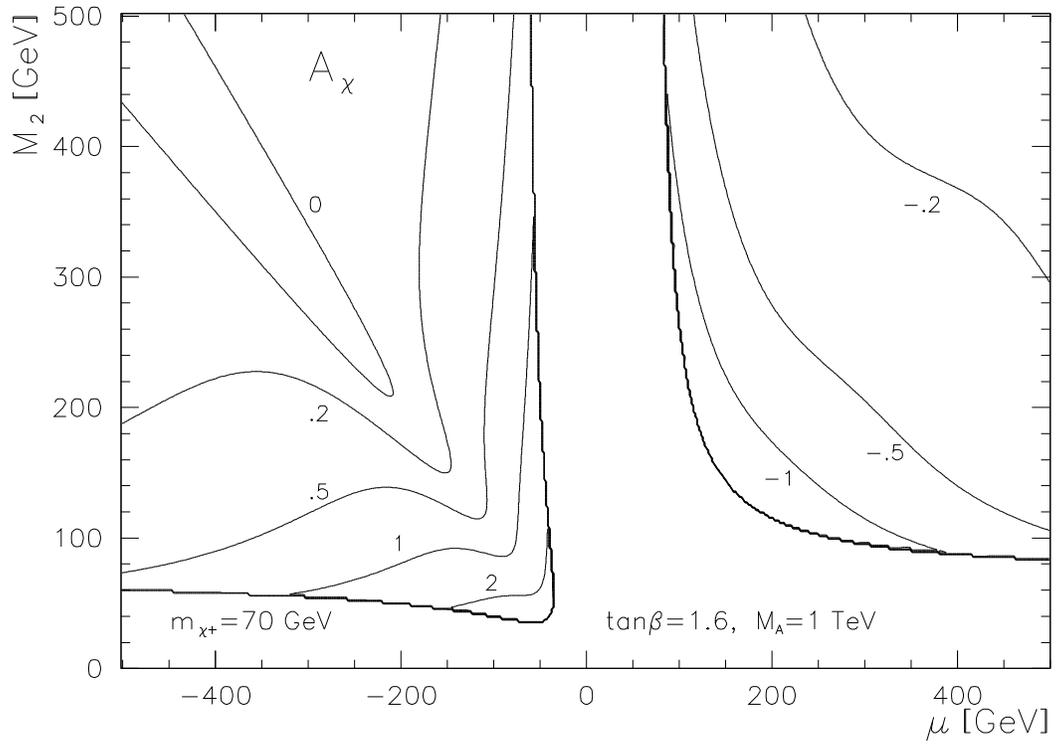
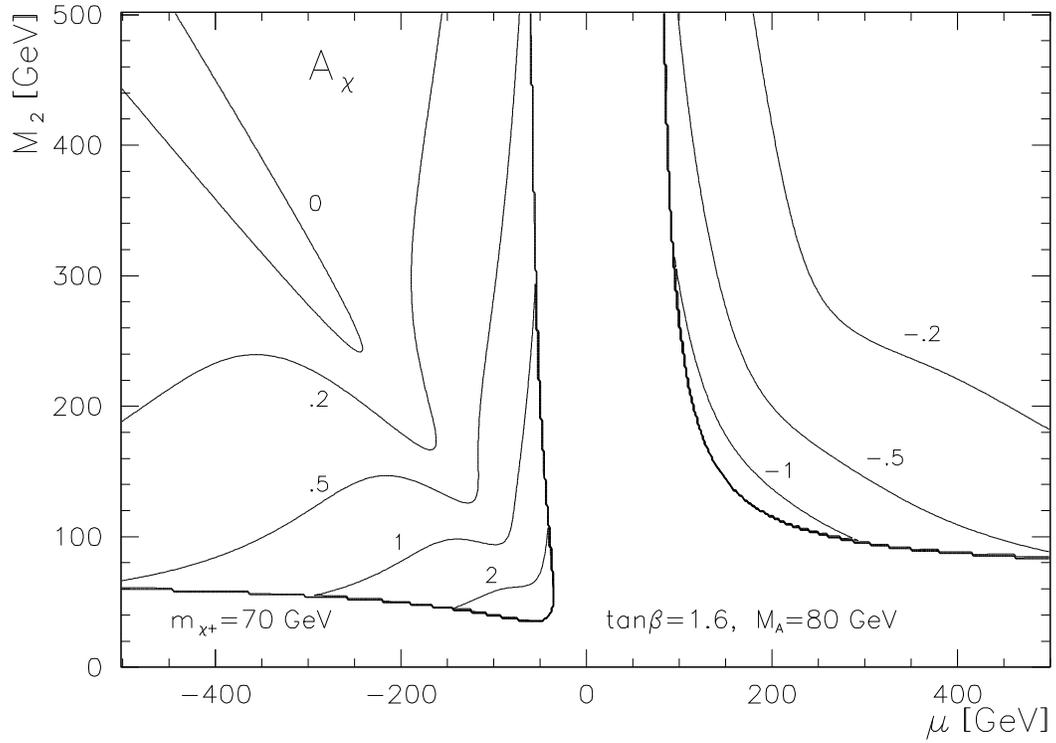
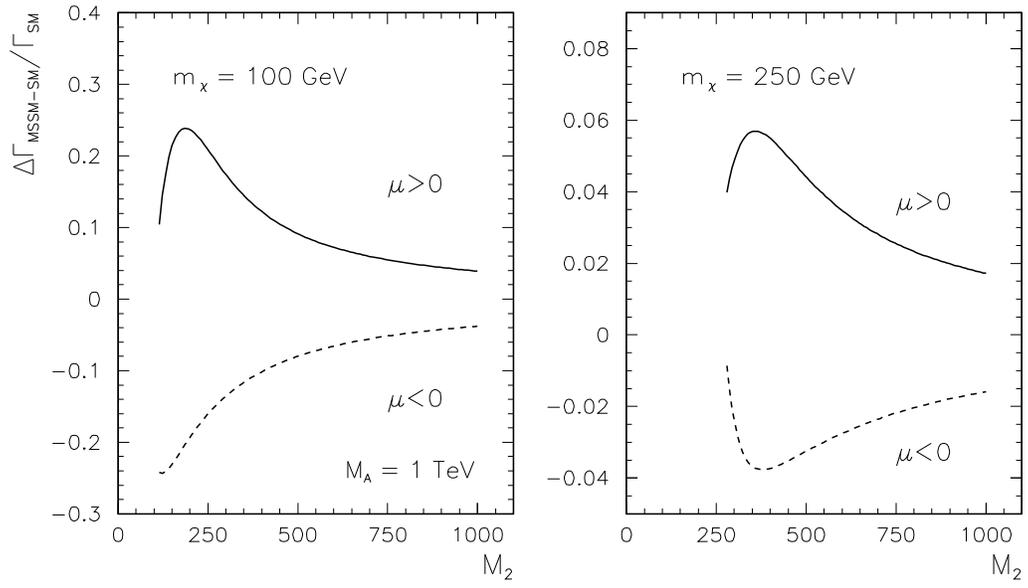
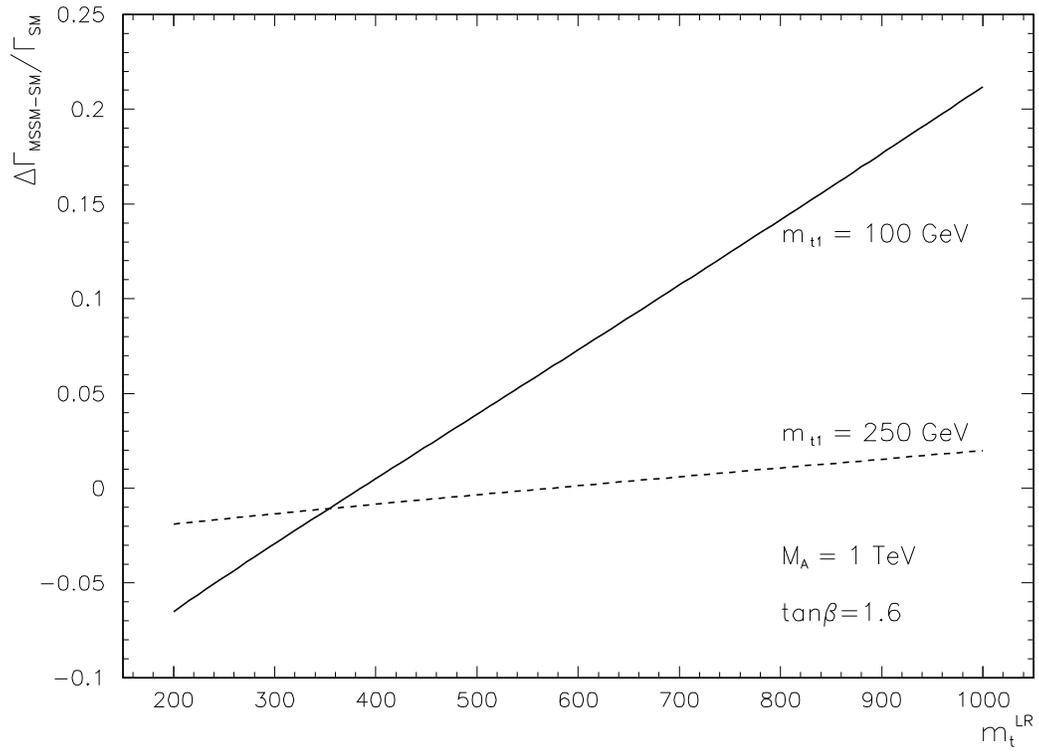
\end{document}